# Small-Worlds: Strong Clustering in Wireless Networks

Matthias R. Brust and Steffen Rothkugel

University of Luxembourg
Faculty of Science, Technology and Communication (FSTC)
6, rue Richard Coudenhove-Kalergi, L-1359 Luxembourg
Luxembourg
{matthias.brust, steffen.rothkugel}@uni.lu

**Abstract.** Small-worlds represent efficient communication networks that obey two distinguishing characteristics: a high clustering coefficient together with a small characteristic path length. This paper focuses on an interesting paradox, that removing links in a network can increase the overall clustering coefficient. Reckful Roaming, as introduced in this paper, is a 2-localized algorithm that takes advantage of this paradox in order to selectively remove superfluous links, this way optimizing the clustering coefficient while still retaining a sufficiently small characteristic path length.

## 1 Introduction

Small-world properties can be found in the World Wide Web, social networks, global economy system, and even in nervous systems [1]. Small-world networks obey two distinguishing characteristics: they have a high clustering coefficient while still retaining a small characteristic path length [2].

Wireless networks belong to the class of spatial graphs, where the links between nodes depend on the radio transmission range, which is a spatial relation between nodes [3, 4].

Small-world properties, namely characteristic path length and clustering coefficient, have a concrete significance related to multi-hop wireless networks. A small path length results in fewer hops what is important for routing mechanisms and the overall communication performance of the entire network [5]. A high clustering coefficient supports local information spreading as well as a decentralized infrastructure. For networks with high clustering coefficient it is supposed that local impacts have a high global effect [6].

In the domain of mobile communication, technologies like for instance Bluetooth and Wi-Fi are able to create communication links within the transmission range at no charge. Additionally, e.g. GMS- or UMTS-adapters can be employed to establish supplementary costly communication links between two arbitrary devices. These links can be established and removed adaptively trying to evoke small-world properties in a spatial wireless network [7, 8]

This work, in particular, concentrates on the optimization of the networks' cluster behavior. For this, a 2-localized algorithm is introduced in order to remove immoderate local links from the network topology. The objective is to create highly clustered regions by removing these superfluous links. The resulting topology is supposed to be considerably more sensitive for additional special long-range links, or shortcuts, and provides a higher potential to exhibit small-world properties.

The remainder of this paper is organized as follows. The subsequent section provides general background on small-worlds. Section 3 describes Reckful Roaming, a 2-localized algorithm for creating highly clustered networks. Experiments and topology properties are discussed in Section 4. Finally, Section 5 concludes this paper.

## 2  Small-Worlds

Many classes of regular networks e.g. square grid networks or the restricted class of circular networks studied in [9] have a high clustering coefficient, i.e. nodes have many mutual neighbors, but large characteristic path length, i.e. large number of hops between pairs of nodes. The formal definition is given below.

**Definition (Local Clustering Coefficient)**. The *local clustering coefficient $CC$* of one node $v$ with $k_v$ neighbors is

$$CC_v = \frac{|E(\Gamma_v)|}{\binom{k_v}{2}} \tag{1}$$

where $|E(\Gamma_v)|$ is the number of links in the neighborhood of $v$ and $\binom{k_v}{2}$ is the number of all possible links.

The clustering coefficient $CC$ then is the average local clustering coefficient for all nodes [10]. The clustering coefficient reflects local characteristics of a network while the characteristic path length expresses global characteristics. The definition for the characteristic path length is as follows.

**Definition (Characteristic Path Length)**. The characteristic path length $CPL$ of a graph is the median of the means of the shortest path length connecting each node $v \in V(N)$ to all other nodes. That is, calculate $d(v,j) \forall j \in V(N)$ and find $\bar{d}_v$ for each $v$. Then define $CPL$ as the median of $\{\bar{d}_v\}$ [10].

The opposite extreme to regular networks are *random* networks, which have a small characteristic path length but exhibit very little clustering. Networks between these two extremes can be constructed by starting with a regular network and moving one endpoint of each edge with a probability $p$ to a randomly chosen node of the network. Thus, the edge remains on the old node with probability $1 - p$. Regular networks correspond to $p = 0$ while random networks are obtained by setting $p = 1$.

Watts and Strogatz [11] discovered that the characteristic path length decreases rapidly as $p$ increases, but the clustering coefficient decreases more slowly. Small-world behavior can then be found in between regular and random networks. They obey a high clustering coefficient as well as a small characteristic path length. It was shown that the initial topology and the construction are not important to arrive at the essential characteristic of a small-world network [9]. Thus, for any given network, it is possible to determine whether or not it is a small-world graph without knowing anything about its construction.

As pointed out, a small-world network must show a specific correlation between characteristic path length and clustering coefficient (small-world properties). There are different equivalent approaches to find this correlation. This work, in particular, uses the following definition [11]. A small-world graph is a graph with $n$ vertices and an average degree $k$ that exhibits a characteristic path length $CPL \approx CPL_{Random}(n,k)$, but a clustering coefficient $CC \gg CC_{Random}(n,k) \approx \frac{n}{k}$. At this, $CPL_{Random}(n,k)$ and $CC_{Random}(n,k)$ are the $CPL$ and $CC$ of a random network with parameters $n$ and $k$.

## 3 Reckful Roaming

The focus in this paper lies on the idea that the clustering coefficient $CC$ can be optimized for spatial wireless multi-hop networks. For this, a topology control algorithm called *Reckful Roaming* (*RR* for short) is introduced as follows.

The definition of the clustering coefficient might induce that more links in each node's neighborhood result in a higher clustering coefficient.

Our approach, however, is based on the observation that this statement does not hold in general. Figure 1 gives an illustrative example.

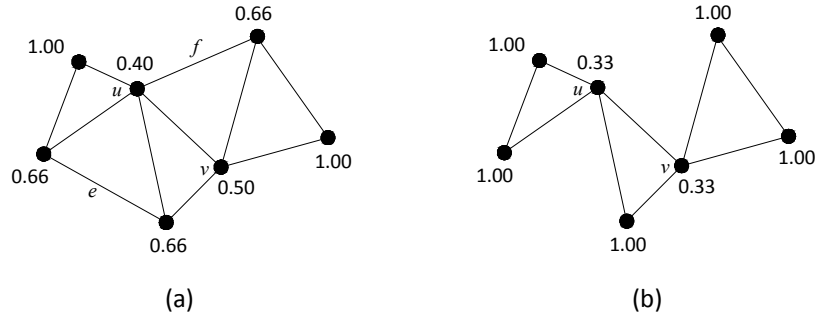

**Figure 1.** Removing links can increase the clustering characteristics of a network.

In Figure 1a nodes are distributed uniformly and connected when two nodes are in spatial neighborhood of each other. That means, two nodes are connected by a link if they are in transmission range. According to this all possible links are created. The local clustering coefficient is denoted adjacent to each node in the figure. The average clustering coefficient is $CC = 0.70$ (cf. Table 1).

In Figure 1b the links $e$ and $f$ are removed. Although the local clustering coefficient $CC_u$ and $CC_v$ for nodes $u$ and $v$ decreased, the average clustering coefficient $CC$, however, increased to 0.81. Values for the characteristic path length $CPL$, average shortest path length $ASP$, average node degree $k$, and network diameter $d$ is given in Table 1.

**Table 1.** Network characteristics for topology illustrated in Fig. 1.

|  | $G_{Init}$ | $G_{RR}$ |
|---|---|---|
| $|N|$ | 7 | 7 |
| $|L|$ | 11 | 9 |
| $CC$ | 0.70 | 0.81 |
| $CPL$ | 1.29 | 1.71 |
| $ASP$ | 1.57 | 1.76 |
| $k$ | 3.14 | 2.75 |
| $d$ | 1.71 | 1.71 |

Reckful Roaming provides a generic approach to implement this general idea of optimizing the clustering coefficient. Increasing the clustering coefficient outperforms the slight increase of the average shortest path length. Moreover, introducing an additional shortcut can easily compensate for that. The basic idea of the algorithm is to verify if a link $(u,v)$ is inefficient in terms of the clustering coefficient. In that case, removing the link $(u,v)$ is taken into consideration (lines 3-8). It is not removed immediately because removing might be advantageous for that particular node only. However, since a link removing affects the local clustering coefficients of the 2-hop neighborhood of the set $\{u,v\}$ an additional remove-confirmation phase will be performed (line 11).

In the confirmation phase, nodes exchange the remove-candidate with that corresponding neighbor. For example, if node $v$ calculates that link $(v,u)$ is a remove-candidate, it sends this result to node $u$. In case that node $u$ has calculated the link $(v,u)$ as remove-candidate as well, the final phase is executed; otherwise the link $(v,u)$ is not removed. Without this optimization stage each remove-candidate would lead to the final phase.

This optimization is justified by the fact that removing a 1-hop link causes a considerably higher impact on the local $CC$-value than removing a 2-hop link [12]. Since a link removal affects the $CC$-values of the 2-hop neighborhood of nodes $\{u,v\}$, the final phase calculates the 2-hop neighborhood of $\{u,v\}$. Both the current $CC$-values of the 2-hop neighborhood as well as the value after a hypothetical removal of $(v,u)$ are calculated. Based on a comparison of those values, it is finally decided whether to remove the link $(v,u)$ (lines 12-18).

As additional criterion, connectivity is guaranteed by line 16 in the pseudo code where removing $(u,v)$ requires that at least one neighbor of $u$ is connected to one neighbor of $v$. Reckful Roaming with the connectivity condition is abbreviated $RR_C$. In case partitioning is a valid option, i.e. when discarding the connectivity condition, the algorithm is denoted $RR$.

Furthermore, note that 2-hop synchronization is required before finally removing a link (cf. line 17-18), since 2-hop topological information is required in order to plan the action. Then again, this local link removal affects the 2-hop neighbors. The synchronization procedure is not detailed here.

An example of applying $RR_C$ on an initial topology and the resulting topology can be found in Figure 2 and Figure 3, respectively.

**Algorithm** Reckful Roaming $RR_C$ (for node $v$)

| | |
|---|---|
| **Input:** | $N$: Initial set of neighbor nodes |
| | $N2$: Neighbors of $u \in N$ (to calc $CC$) |
| | $RC$: Set of remove candidates |
| **Output:** | $N_{RR}$: Resulting neighbor nodes |

```
01:    CC(v) ← |E(Γ_v)|/(k_v choose 2)
02:    RC(v) ← ∅
03:    for each u ∈ N do
04:        N(v) ← N(v) \ {u}
05:        CC_rc(v) ← |E(Γ_v)|/(k_v choose 2)
06:        if CC_rc(v) > CC(v) then
07:            RC(v) ← RC(v) ∪ {u}
08:        N(v) ← N(v) ∪ {u}
09:    sendRC(v, RC(v))
10:    receiveRC(u, RC(u))
11:    for each w ∈ (RC(v) ∩ RC(u)) do
12:        CC(uv) ← Σ_{x ∈ (N(v) ∪ N(u))} CC(x)
13:        N(v) ← N(v) \ {w}
14:        CC_rc(uv) ← Σ_{x ∈ (N(v) ∪ N(u))} CC(x)
15:        if CC_rc(uv) > CC(uv) then
16:            if N(v) ∩ N(u) = ∅ then
17:                if permit(v) = true then
18:                    N(v) ← N(v) ∪ {w}
19:    N_RR(u) ← N(v)
```

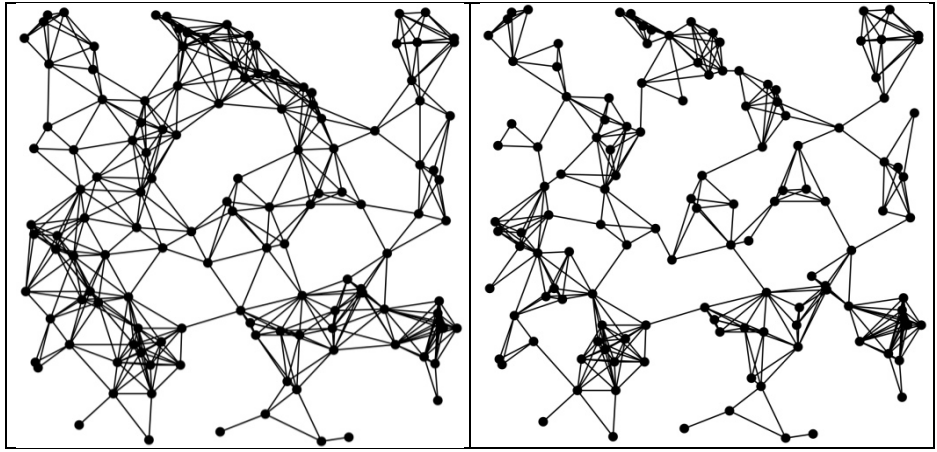

**Figure 2.** Initial network with $k^* = 9.96$ (left) and final topology after applying $RR_C$.

## 4 Empirical Study

### 4.1 Experimental Setting: Network Density

Transmission range, simulation area, and number of devices are typical simulation parameters. We introduced the notion of network density in order to identify classes of network topologies with similar characteristics. This way, results obtained for some particular network density apply to a whole family of networks.

**Definition (Network Density).** The *network density* $k^*$ is defined as

$$k^* = \frac{1}{l^2} \sum_{v \in N} Cov\big(RA(v)\big) \qquad (2)$$

that is $k^*$ is the sum of the total coverage area $Cov$ of all nodes $v \in N$ divided by the simulation area $l^2$. The transmission range is given by a range assignment function $RA$.

Thus, the network density can be increased by either increasing the transmission range or the number of devices as well as by decreasing the simulation area.

Often the network density of a wireless network is defined by the number of devices per square unit [13]. However, this definition returns the same value for the same network configuration with different transmission ranges.

We assume a square area as simulation area as well as circular coverage areas. In case where all devices have the same circular transmission range the formula for network density can be written as $k^* = \frac{\pi \cdot tr^2 \cdot n}{l^2}$.

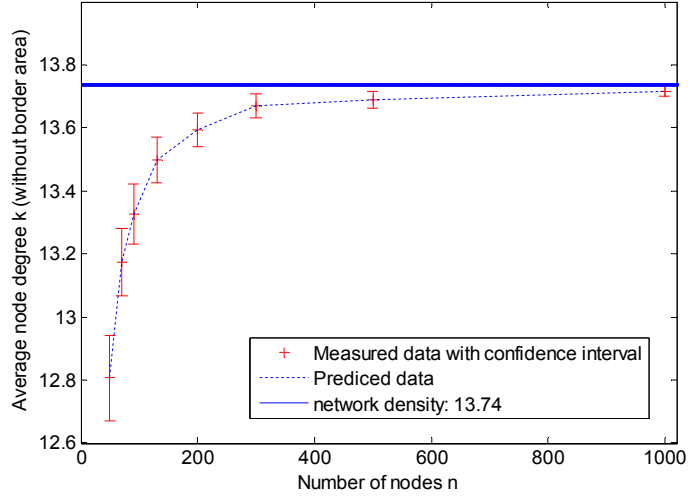

**Figure 4.** Node degree $k$ without border area for $k^* = 13.74$. Confidence level of means is set to 95%.

Figure 4 indicates that the average node degree of considered nodes converges to $k^* = 13.74$ when increasing the number of nodes. We deduce that the network density $k^*$ represents the asymptotic value for the average node degree of a network (when ignoring nodes in the border area).

In order to further validate that, several additional experiments with different network densities have been conducted. Table 2 fosters our assumption for different network densities.

**Table 2.** Empirical data for $|N| = 1000$.

| $k^*$ | $\mu$ | $ci$ (95%) | $\sigma$ |
|---|---|---|---|
| 6.11 | 6.10 | ±0.0087 | 0.14 |
| 9.96 | 9.95 | ±0.0127 | 0.20 |
| 13.74 | 13.72 | ±0.0172 | 0.28 |
| 16.35 | 16.32 | ±0.0201 | 0.32 |
| 25.13 | 25.08 | ±0.0323 | 0.52 |
| 39.27 | 39.17 | ±0.0540 | 0.87 |

Note there might be a more appropriate definition of network density for different settings as those assumed here, in particular when considering elliptical or non-uniform coverage areas.

Subsequently, the correlation between the network density and the average node degree $k$ is analyzed.

Figure 4 shows the setting of the experiment. A network density of $k^* = 13.74$ is used. The number of nodes has been varied from 50 to 1000 while adjusting the transmission range and the simulation area such that the network density of 13.74 remains constant. The average node degree has been measured for each setting while nodes in the border area have been ignored, because the border area of $\frac{tr}{2}$ length from each side includes nodes that are obviously less connected to other nodes.

### 4.2 Experiments

In our experiments, fully connected topologies are considered. A fully connected input topology for $RR_C$ results in a connected output topology. For a reasonable setting, what network density is a lower limit for a connected network with high probability? Initial experiments (not stated here) indicate that a network density higher than six result in a connected network in most cases.

Interestingly, this corresponds to the "magic number" found by Blough for weak connectivity [14]. In the general form, however, there is no constant for the minimum number of neighbors for full connectivity as proven by [15].

Experiments were conducted for network densities $k^* = 6.11$, 9.95, 13.7, and 25.13. The clustering coefficient $CC$, average node degree $k$, characteristic path length $CPL$, network diameter $d$, and the critical transmission range $CTR$ were analyzed. The critical transmission range of a static wireless network represents the minimum transmission range where the network remains still fully connected.

Since $RR$ is not aimed at optimizing the $CTR$ we do not expect any improvement on the final topology, but included this metric for reasons of completeness. For each network density, sufficient runs have been conducted arriving at a confidence level of 95% for all measurements. Since $RR$ is not likely to stabilize when applied a single time only, we executed $RR_C$ four times for each topology.

The results are shown in Figures 5 to 8. All values have been normalized using the initial topology values as normalization values. The first phase represents the initial topology before applying $RR_C$ for the first time.

### 4.3 Results and Properties

Figures 5 to 8 reveal that the clustering coefficient increases between approx. 8% for $k^* = 6.11$ and 19% for $k^* = 25.13$. The average node degree decreases between approx. 20% for $k^* = 6.11$ and 42% for $k^* = 25.13$. While $k$ is decreasing the augmentation of the characteristic path length starts slower with approx. 7% for $k^* = 6.11$, but increases faster, reaching around 43% for $k^* = 25.13$. The network diameter $d$ increases proportionally to the characteristic path length. As expected the $CTR$ does not decrease considerably.

We deduce that following correlations exist. The clustering coefficient appears to be proportional to characteristic path length as well as network diameter, thus $CC \sim CPL$ and $CC \sim d$. Furthermore, the characteristic path length seems to be in inverse proportion to the average node degree $CPL \sim \frac{1}{k}$.

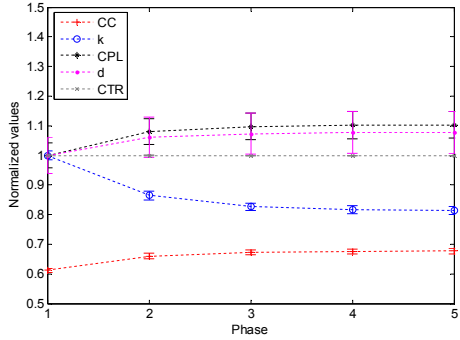
**Figure 5.** $RR_C$ results for $k^* = 6.11$.

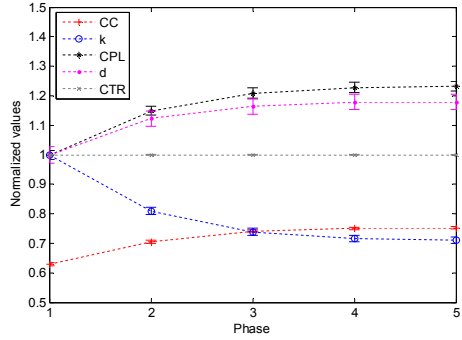
**Figure 6.** $RR_C$ results for $k^* = 9.96$.

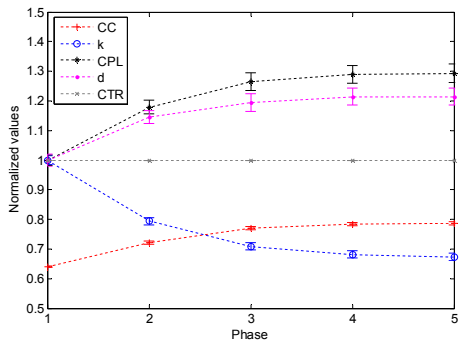
**Figure 7.** $RR_C$ results for $k^* = 13.75$.

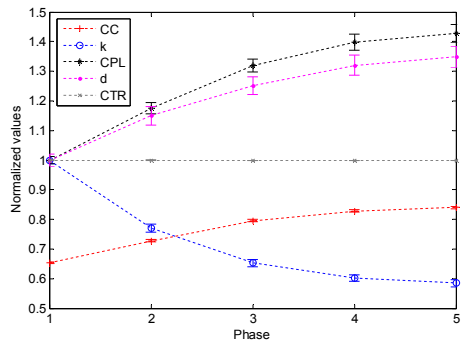
**Figure 8.** $RR_C$ results for $k^* = 25.13$.

The main effect of $RR_c$ and $RR$ is to optimize the clustering coefficient. This is facilitated by selectively removing links. Link removals are more likely to occur in sparse regions, while highly clustered regions are mostly unaffected. The resulting topology is connected in case of $RR_C$, while in case of $RR$ it is not necessarily connected.

$RR$ does not fully optimize the topology in a single run. However, our experiments have indicated that after approximately three runs, only negligible further optimizations can be obtained. After four to six runs, the values finally stabilize.

The resulting topology depends on the order $RR$ is executed on the nodes. Simulations, however, have shown that the final $CC$-values do not differ considerably (estimating approx. 1%).

## 5  Conclusions

Small-world networks obey two distinguishing characteristics: they have a high clustering coefficient while still retaining a small characteristic path length. These small-world properties induce several benefits. Whereas most approaches focus on augmenting the network by well-chosen shortcuts in order to evoke small-world properties, our approach focuses on creating highly clustered regions by removing inefficient links in terms of the clustering coefficient.

For this purpose, a 2-localized algorithm called Reckful Roaming (*RR*) is described. Reckful Roaming preserves connectivity in the worst case ($RR_C$). This work showed that *RR* removes clustering-inefficient links and considerably increases the clustering coefficient. This is an initial yet major step to foster evoking small-worlds in spatial networks like sensor networks.

A highly clustered network is even more sensitive to well-placed shortcuts, so that we expect an even more beneficial effect when applying these shortcuts to *RR*-generated topology.